\documentclass[conference]{IEEEtran}

\usepackage{cite}
\usepackage{color}
\usepackage{multirow}
\usepackage{booktabs,multirow,array}
\usepackage[table]{xcolor} 
\usepackage{ltxcmds}
\usepackage{footnote}
\usepackage{amsthm}

\usepackage[table]{xcolor}
\usepackage{subfigure}
\usepackage{lipsum}
\usepackage{amssymb} 
\usepackage{pdflscape}
\usepackage{afterpage}

\ifCLASSINFOpdf
\usepackage[pdftex]{graphicx}
\else
  \usepackage[dvips]{graphicx}
\fi

\begin{document}
\title{Controlling Resource Allocation using Blockchain-Based Delegation}

\author{\IEEEauthorblockN{Shantanu Pal$^{\ast}$, Ambrose Hill$^{**}$, Tahiry Rabehaja$^{\mathsection}$, Michael Hitchens$^{\mathsection\mathsection}$}
\IEEEauthorblockA{$^{\ast} $School of Computer Science, Queensland University of Technology, Brisbane, QLD 4000, Australia\\ $^{**}$School of Computer Science and Engineering, University of New South Wales, Sydney, NSW 2052, Australia \\ $^{\mathsection}$Risk Frontiers, Sydney, NSW 2065, Australia \\ $^{\mathsection\mathsection}$Department of Computing, Macquarie University, Sydney, NSW 2109, Australia
\\
{shantanu.pal@qut.edu.au, ambrose.hill@data61.csiro.au, tahiry.rabehaja@riskfrontiers.com, michael.hitchens@mq.edu.au} {}}}




\maketitle


\begin{abstract}
Allocation of resources and their control over multiple organisations is challenging. This is especially true for a large-scale and dynamic system like the Internet of Things (IoT). One of the core issues in such a system is the provision of secure access control. In particular, transfer of access rights from one entity to another in a secure, flexible and fine-grained manner. In this paper, we present a multi-organisational delegation framework using blockchain. Our framework takes advantage of blockchain smart contracts to define the interactions and resource allocation between the consortium of organisations. We show the feasibility of our solution in a real-world scenario using the allocation of transportation credits in a multi-level organisational setting as a use-case.  We provide proof of implementation of the proposed framework using the Hyperledger Fabric blockchain platform. Our results indicate that the proposed framework is efficient and can be used for city-wide transport, potentially even scale country-wide with a shared blockchain with complex access control rules. It also bestows better transparency to the delegation of access rights and control over the employees' transportation access for the organisations. 
\end{abstract}

\IEEEpeerreviewmaketitle

\section{Introduction}
\label{introduction}
In an increasingly digitised and decentralised environment, groups of users will wish to access, employ and expend resources in ways that are both co-operative and dynamic. Groups of users will draw on common resources (e.g., digital currency), to which they have authorised access~\cite{magiera2005security}. However, even with pre-existing authorisation, the extent and nature of the use they make of the resource may not be known in detail in advance. Limiting each user to an average portion of the resource may not be an efficient use of the resource, as some users will wish for greater than average access, some for less than average.  Allowing each user access to a greater than average share of the resource risks over use. For example, a group of users, may be able to access 100 units of a resource between them. Say there are 10 members of the group.  An average allowance would be 10 units. However, if not all users access this much, some of the resource goes to waste.  Allowing all users access to more than 10 units risks over-allocation. While, in centralised systems, this situation can be easily handled, in distributed environments, where resources are supplied by multiple providers and accessed by users external to that provider, ensuring efficient access without over-committing the resource is a more complex issue~\cite{pal2019design}. In such situations, the providers and consumers of resources may not want to be at a disadvantage in terms of the control of the relationship. That is, they may want the relationship supported by an infrastructure that treats all participants equally.

In situations such as these the issue of access control delegation is significant~\cite{pal2021internet}. In such an environment, multiple attributes possessed by an entity need to be validated by multiple different authorities that regulate the access control policies. Delegation denotes the transfer of access rights from one entity to another that is governed by a set of access control policies and associated conditions to access a resource~\cite{4426056}. That said, the delegation of access rights determines the level of access that must be transferred from one entity to another, governed by the set of access control policies and associated access conditions.  In a closed environment, such delegation may be controlled by a central entity employed to perform this function. But when the system grows, for example, in a multi-organisational context, the multi-level delegation needs to be performed over a distributed system spanning across multiple jurisdictions. Given the scale, dynamic nature, and massive network complexity of interconnected devices, users, applications, and services (e.g., in an Internet of Things (IoT) setting \cite{atzori:iot-survey-2010}), this becomes a challenging task, in particular, to control resources allowing access only to the authorised users. 

\subsection{Motivation and Problem Statement}
\label{motivation-use-case}
Consider the challenge of shared resources discussed above. Suppose an organisation wants to create an automated credit system to maintain and manage costs associated with the transportation-related expenses of employees. This is achieved through the use of a pool that provides flexibility and efficiency in credit management. Employees access the credit pool through delegation which has its root in access control systems. The delegated access to the pool can then be attached with conditions and tree-like hierarchies which provides an expressive and flexible platform upon which granular and complex rules can be written.
Through delegation, employees have access to the credit pool under given and specific conditions that are automatically verified. For our purpose, we call this a \textit{transportation credit}. Transportation credit can be seen as an amount of money (virtual) that an employee can spend in a certain period of time to perform some specific tasks. For instance, an employee can travel by bus, train, or flight to join a business meeting, perform remote operational tasks or simply commute in-between work sites. In general, the conditions can use spatio-temporal properties as well as other aspects such as employee ranks or mode of transportation. The company purchases an amount of transportation credit and allocates access to employees where the sum of the access can exceed total purchased. This allows flexibility to meet individual work requirements while still maintaining an overall budget.

 \begin{figure}[t]
 \centering
    \includegraphics[scale=1.35]{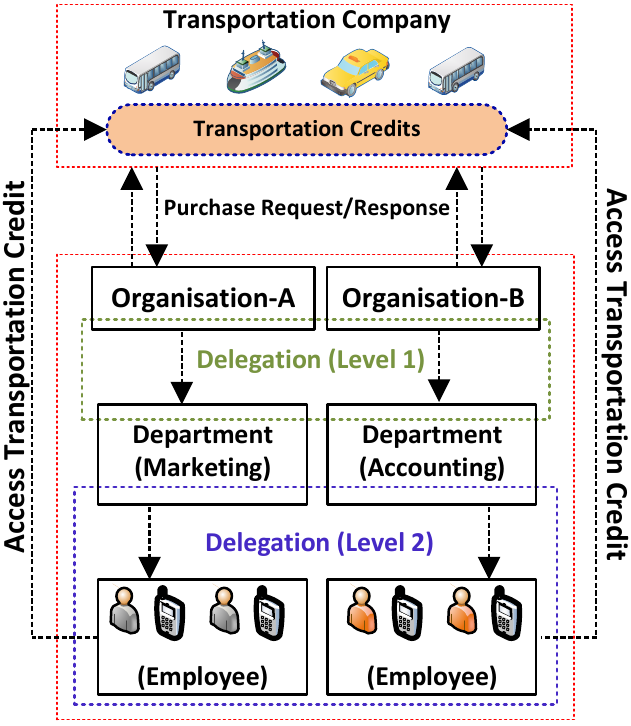}
    \caption{A conceptual view of the delegation process in multi-organisations.}
    \label{fig:ac-1}
    \small
 \end{figure}

The organisation contacts a preferred transport company (or several transport companies) to buy some transportation credit (in exchange for crypto or fiat currency). After evaluation of the request and successful payment, the transportation company sends the appropriate credit to the organisation. Upon receipt, the organisation allocates the credits based on the access control policies and enforceable conditions. The access rights are delegated to the intended departments, and the department further delegates access rights (with more conditions, so as to ensure monotonicity) to the employees working within that department. In Fig.~\ref{fig:ac-1}, we illustrate a conceptual view of such a delegation process. Note, there may be multiple organisations and multiple transport companies present in the system. There is a `channel' between the organisations and transportation companies to mutually agree on the services and the associated costs. That is, the transportation credit has some top-level access control between the organisation and consortium of transportation companies (different companies managing the various transportation services). For each organisation, there may be many employees working in various departments. Effectively, each organisation can purchase a certain amount of credit, store them in a `pool' (controlled by the organisation), and then allow employees to access them on demand, as long as the access control rules are satisfied. Employees, once their company has joined the network and purchased some amount of transportation credit, can then request to use, and access these credits in near-real-time. Access to these transportation credits (i.e., resources) would be, for example, via a mobile application that will allow the employees to request, check and use the credits. Note, we use the term company to show the business, and the term organisation to show a large-scale business. In essence, they can be interchangeable, and we use these two separate terms for ease of reading. 

The example just described is an example of the broader situation described above, that is, one where we may have an overall budget, and multiple entities that can access that budget, but where a precise pre-allocation of resources is non-optimal.
To provide fine-grained access control delegation for controlling resource allocation, several mechanisms can be used~\cite{ravidas2019access}~\cite{pal2017towards}. For example, Role Based Access Control (RBAC) can provide fine-grained access control over the resources explicitly using user-to-role mappings. However, in RBAC, for every access to a resource, there is a need to define separate user-to-permission relations. Moreover, RBAC is highly centralised in nature, limiting its scalability for large-scale dynamic systems. Attribute Based Access Control (ABAC) uses attributes to improve the policy management rather than depending upon the concrete unique identity of individual entities. However, ABAC does not, for example, support mechanisms for grouping together policies with different attribute requirements that allow access to a single resource or policies with the same attribute requirements that allow access to different resources. That said, ABAC requires a policy management mechanism, especially when the number of policies rises significantly~\cite{pal2018policy}. Capability Based Access Control (CapBAC) provides flexible access control by distributing capability tokens that contain access rights or privileges (along with additional conditions) validated at the edge IoT devices at the time of access to a resource. However, most of the CapBAC systems are centralised for policy storing and their management~\cite{pal2019policy}. Therefore, these access control mechanisms can not provide scalable, robust, and trustworthy delegation supporting the decentralisation of resources. 

We argue that managing such resource delegation in a multi-level scenario involving various dynamic and large-scale systems requires scalable, efficient, lightweight, trustworthy, and robust policy enforcement mechanisms~\cite{rabehaja2019design}. The delegation process should scale with the system through decentralisation. At the same time, there is a need to establish trust between the various entities within the system to delegate access rights~\cite{pal2019integration}. 

Blockchains have the potential to address these issues and overcome the limitations of traditional access control mechanisms in a more efficient and fine-grained way for large-scale IoT systems spanning multiple jurisdictions \cite{koh2020blockchain}. The distributed platform of blockchain allows an apparent balance of power where no participating entity has sole control of the system. This means that agreement must be achieved by parties regarding any addition, deletion, or update of contract and payments for every transaction. This creates a balanced atmosphere for the clients and service suppliers through an open marketplace platform where supply and demand are dynamic. In addition, using blockchain as a base delivers new opportunities by providing distributed storage and computational framework on which programs can be executed, which ultimately leads to efficiency and automation. Several properties of the blockchain (e.g., no central authority and trusted third-party, consensus mechanism, immutable, irreversible and tamper-proof, accessibility, auditability, etc.) offer secure and safe foundations to record and track a list of transactions for a large number of devices in a highly transparent, auditable and efficient way~\cite{pournader2020blockchain}. Moreover, our use of Hyperledger enables the segmentation of the market into channels where private interactions between an organisation and multiple transport suppliers is supported on top of the underlying global blockchain. This means that, even though the native payment system is universal to all parties, secure and private interactions can be established dynamically while inheriting the aforementioned blockchain properties.


\subsection{Contributions}



We employ blockchain to provide granular and fine-grained decentralised management of access control delegation to resources within a multi-organisational establishment. This includes an entity-centric delegation tailored to various system granularities spanning over multiple timescales. The main contributions of the paper can be summarised as follows: 

\begin{itemize}
    \item We propose a framework that can manage resources owned by a consortium of entities, and assists allocation of these resources through access control delegation. 
    \item We present the system design along with a detailed framework of the proposed delegation model. We develop a distributed and decentralised delegation framework for fine-grained access rights transfer using blockchain.
    \item We provide a detailed proof of concept prototype implementation in the Hyperledger Fabric blockchain platform to show the feasibility of our system. 
\end{itemize}

The rest of the paper is organised as follows. In Section~\ref{delegation-framework}, we discuss system components, system functionality, and communication protocol of the proposed delegation framework. In Section~\ref{implementation-evaluation}, we present system implementation and evaluation of the achieved results. Section~\ref{related-work} includes related works. Finally, we conclude the paper with direction to the future work in Section~\ref{conclusion}.  

\section{Proposed Delegation Framework}
\label{delegation-framework}
In this section, we discuss (i) the various components of the proposed delegation framework, (ii) the system functionality, and (ii) show the communication protocol between the various components of the framework. 

\subsection{Components}
\begin{figure}[t]
 \centering
    \includegraphics[scale=1.65]{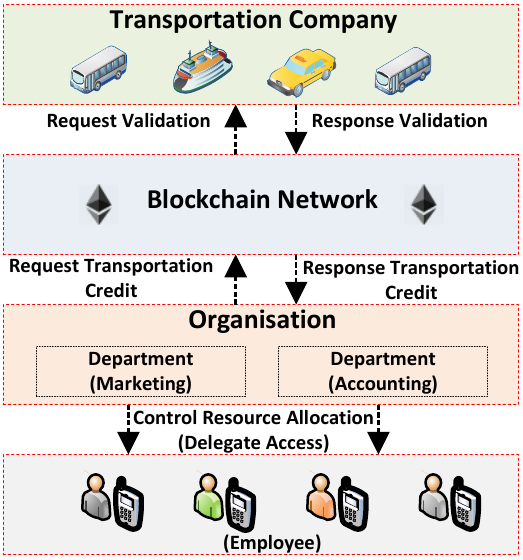}
    \caption{Layer view of the proposed blockchain-based delegation framework.}
    \label{fig:ac-2}
    \small
 \end{figure}

In Fig.~\ref{fig:ac-2}, we illustrate the proposed framework. It is composed of the following major components: 

\begin{itemize}
    \item \textit{Employee}: Who works within the company and wants to access transportation credit. We assume that all employees have a smart device that can be used to access a resource, as well as able to perform the communication between the various entities within the organisation. 
    
   \item \textit{Organisation}: Where a group of people works together for a particular purpose in pursuing dedicated objectives to e.g., a business, institution, or an association. There may be various departments within a company within which groups of people function. 
   
   \item \textit{Blockchain Network}: A dedicated distributed infrastructure spanning over the entire network to provide ledger and smart-contract services autonomously. We select a private blockchain network as all participating organisations can be identified, and to restrict access only to members of those organisations. 
   
   \item \textit{Transportation Company}: A dedicated infrastructure (or business) that provides transportation services to move people or goods e.g., bus, train, taxi, or flights. It also provides a chain of services from scheduling timetables, managing ticketing, and control the logistics process. 
\end{itemize}

\subsection{System Functionality}
The blockchain network interfaces between businesses wishing to organise transportation credits for their employees, and the transportation companies that facilitate those solutions. We call the overall network a transport chain. We then setup private channels on the transport chain for each participating organisation. A channel here is another distributed ledger that sits on top of the larger blockchain but is only maintained by a subset of the participants. For this use case each organisation would create their own channel and include any transport companies they want to conduct business with. With this design, privacy is guaranteed between organisations.  
Transport companies uses a token smart contract on the transport chain to sell transportation credits (maybe in terms of tokens, tickets, and passes) to the organisations. As noted above, the currency used can be any crypto or fiat currency, 
however, the participants must agree on the currency. Using a crypto currency would allow the financial transactions to take place on the transport chain via the smart contracts. Transport companies would join channels with any organisation looking to purchase transportation credits for that transport type. 

The organisation would also use the smart contract to purchase the transportation credits. However, once they have been purchased the tokens/tickets/etc. can be accessed through access control rules defined by another smart contract deployed and controlled by the organisation. Employees, once their company has joined the transport chain network and purchased some amount of transportation credits, can then request to use and access these tokens in near-real-time. They would have access to the transport chain via a mobile app that will allow them to request, check and use these credits. Note, when an employee requests transportation credits, the \textit{access control smart contract} can automatically check the access request against the access rules. The rules can be based on many metrics including amount, time, role, location, or transport type. 

In other words, when an employee requests access to transportation credits via the organisation's access control smart contract -- the access request validity and whether the organisation has enough credits is checked and stored on blockchain network via a transaction to the token smart contract. The employee can then use the credits at the point of transport, which once used will send confirmation back to the token smart contract, finalising the trip and using up the required amount of credits. The reason two smart contracts are used is so that the organisations have control and privacy over the access control rules but they are still on the blockchain to automate the checking process. Transport companies can guarantee that employees have the right to use credits without seeing what made them eligible.

\subsection{Communication}

In Fig.~\ref{fig:commDiagram}, we illustrate the steps each entity in our system takes to communicate with and use the system. To give a more fine-grained view of the interaction, it is split up into three parts, (i) purchase transportation credit, (ii) access control setup, and (iii) use transportation credit. These could be generalised in the obvious fashion if a different resource was being managed. A detailed description for each of them is as follows: 

\begin{itemize}
    \item Step 1: The organisation and the transportation company will negotiate a price for transport credit (in terms of tokens) that can be used with that particular transportation company. This happens off-chain (i.e., outside the blockchain) and there is no time limit as the deal will need to be in the best interest of both parties. 
    \item Step 2a: Once a price has been agreed upon by both parties, then the blockchain is updated. First, the transportation company will initialise a token smart contract with the price and submit a transaction that places the agreed amount of tokens in escrow (i.e., a financial and legal agreement).
    \item Step 2b: The organisation will send payment, either via another token or off-chain with a receipt being put on-chain, to the smart contract.
    \item Step 3: The contract will wait for both the tokens and payment before releasing the tokens. There is also a check here to ensure that the amount of tokens and amount paid matches the proposal. This is done as there is an initialisation step to the contract that requires the parameters of the proposal to be entered. If any of the transactions do not match the proposal the transaction is rolled back and a new contract will have to be deployed.
    \item Step 4  (a and b): Once the payment and tokens have been received the smart contract will emit confirmation events on the channel notifying the transportation company and organisation that the transaction is successful.
    
\begin{figure}[t]
     \centering
     \includegraphics[scale=1.1]{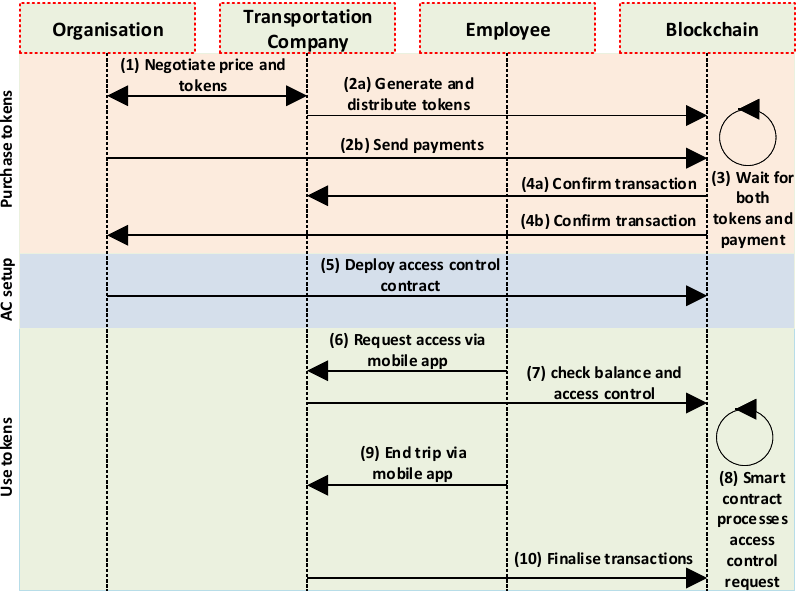}
     \caption{Communications among the different components of the framework.}
     \label{fig:commDiagram}
\end{figure}
    
    \item Step 5: The organisation deploys an access control smart contract on the channel. This contract will have the codified rules for accessing the tokens. This is a completely internal process and the organisation has complete control over which employees can access the tokens, how much they can access for certain services, and even delegate this access control.
    \item Step 6: When an employee wants to access those tokens i.e., use the transport service, they will put in their transport details (start/end destination, transport type, etc.) into their mobile app. This can be done in advance or at the point of interaction with the transportation company.
    \item Step 7: The system (managed by the transportation company) will then initiate a transaction that will take confirm the trip and send it to the access control contract. 
    \item Step 8: The smart contract will process the request by validating two values. Firstly, it will check that the organisation has enough balance for the transport, this could be calculated based on the final destination or just having a maximum spending limit. It then checks that the employee passes the access control rules. If both checks are passed, an event is emitted to the transportation company to signal that the trip can take place. 
    \item Step 9: At the end of the trip, the employee finalises the trip on their mobile app.
    \item Step 10: The transportation company will also send a transaction when the trip is complete which will allow the smart contract to automatically use up the tokens. 
\end{itemize}

\section{System Design and Evaluation}
\label{implementation-evaluation}
In this section, we discuss (i) the implementation details of the proposed delegation framework, and (ii) evaluate the achieved results.

\subsection{Implementation}
We use Hyperledger Fabric (HLF) to implement the proof of concept of our system~\cite{androulaki2018hyperledger}.
HLF is an enterprise-ready blockchain  platform that contains many of the required features (e.g., performance, scalability, and levels of trust) that our system will use. Version 2.2 of HLF is used, and version 0.4.2 of Hyperledger Caliper is used to benchmark the system. The tests are run on a Dell Latitude 7490 Notebook with 8 GB of memory and an Intel Core i5 processor. The smart contracts are written in GO version 1.15.2.

Note, HLF's inbuilt channel system is used to set up each organisation on their own private channel that transport companies can then be added to for price negotiation and token generation. Having separate channels for each organisation allows for private negotiation, the delegation of contracts and usage whilst still storing all key transactions into the blockchain state. The channel system can be visualised as a ledger on a ledger with access control tools placed on top. Every entity will have access to the foundation ledger and that will store information that is public or vital for users to interact with the system, while any of the private information will be stored on one of the smaller ledgers, and only authorised entities will be given access to these ledgers.

As we discussed in the previous section, the implementation consists of two smart contracts, the \textit{token smart contract}, and the \textit{access control smart contract}. There will be one access control contract per channel which the organisation controls. This smart contract is where the organisation's admin can specify rules (and conditions) for the delegation and usage of the tokens on this channel. Each transport company that wishes to interact with this organisation must deploy a token contract which can be done before or after the price negotiations. The price and amount of tokens for each transportation mode can be maintained using this contract. There would be one channel per organisation on the network in a production deployment, with each channel having one access control contract and as many token contracts as transport companies on that channel. 

Our test network consists of two organisations and two transport companies, each on its own channel. The transactions are generated in workloads using caliper. To reduce strain on the test machine, the two contracts were combined and deployed together. For testing purposes, two fewer docker containers are needed, and it functions theoretically the same as though two contracts were deployed. The simulation was run ten times, each time simulating over 20,000 transactions. The transactions simulated were as follows:
\begin{itemize}
    \item Request\_Access: The transaction a user would send when trying to use a transport token. The smart contract will check that this user has permission to use tokens and that the organisation has the tokens to be used. 
    \item Finish\_Trip: The transaction a transport company would send when it detects a user has completed their trip using this service. The trip object in the blockchain state is updated, saying it has been complete and any tokens associated with that trip are now considered spent. 
\end{itemize}

Each round of testing involved simulating different send rates for the system, starting at 100 transactions per second (tps) and ending at 300 tps, increasing the send rate by 50 tps each time.

\subsection{Results and Discussion}

\begin{figure}
     \centering
     \includegraphics[scale=.43]{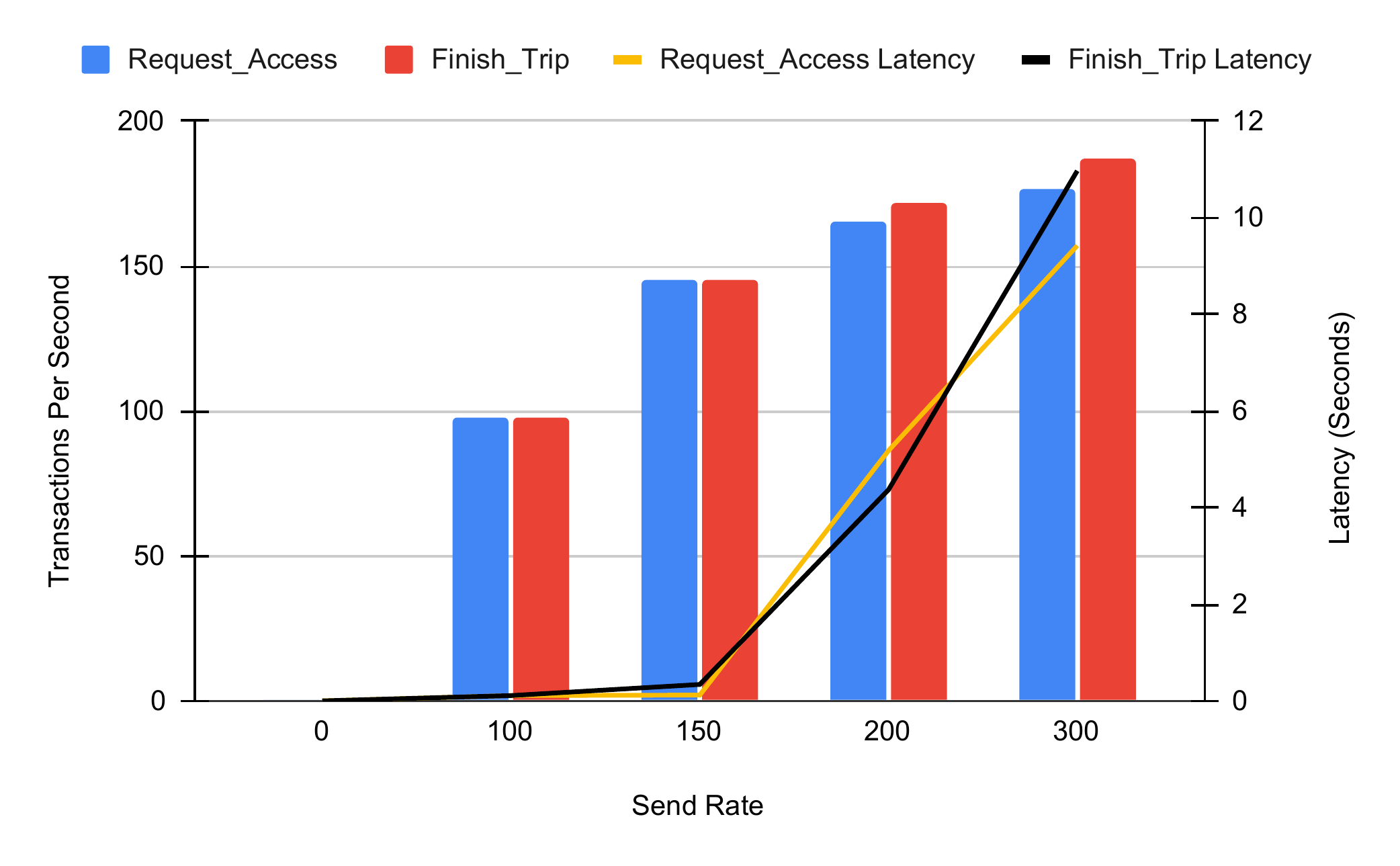}
     \caption{Studying the throughput and latency of the proposed framework.}
     \label{fig:results}
\end{figure}


In Fig.~\ref{fig:results}, we illustrate the achieved results of the average throughput and latency of the system. Next, we discuss the throughput, and then the latency in detail. We also discuss the suitability of the achieved results of the proposed system in practical scenarios. 

We observe that the system reaches throughput saturation between 175 and 200 transactions per second. \textit{Request\_Access} is the more computationally expensive transaction. It is expected to have a smaller throughput than the \textit{Finish\_Trip} transaction, which can be seen when the send rate is high (i.e., more than 200 tps), and the system is past saturation. Below saturation, the system is able to process these transactions similarly. 
With the latency results, it is clear to see that the latency is trivial before the system reaches saturation as it is constantly less than 0.5 seconds. When the system does reach saturation, there is an upwards trend in latency. However, this is expected as when the system is under stress from many transactions, for each second there are more transactions than the system can handle, the backlog of pending transactions continues to grow, leading to longer wait times for confirmed transactions. At a workload of 300 transactions per second, the latency is seen to be 
10 seconds for both transactions. 

With a throughput of 175 tps and a maximum latency of 10 seconds, the prototype system is more than capable of handling a transport application of a large size (i.e.,  many concurrent users). Transactions such as funding and token generation can be processed in advance and so do not need to be taken into account when benchmarking the system. The critical transactions, such as requesting access, need to be handled as efficiently as possible as they will be executed at the point of interaction between user and transportation mode. 
It can be noted that depending on the size of the organisations and transport companies, our system would be able to handle a large number of users all requesting access and finishing trips at the same time. Moreover, if the situation does occur where there are more than 175 requests a second, a user would not be waiting more than 10 seconds for confirmation.

These results indicate that this system could be used for city-wide transport, potentially even country-wide, with a sharded blockchain. Even organisations and transport companies with complex access control rules will experience short wait times, transparency into transport usage, trusted digital history of use, and total control over their employees' transport access can easily be achieved to a flexible and fine-grained manner. 


\section{Related Work}
\label{related-work}
Recently, there has been significant attention on blockchain and smart contract technology for authentication, access control and delegation of access rights \cite{pal2017design-mobiquitous} \cite{pal2017design-icst}. Several proposals have discussed the need for blockchain technology in transport and logistics to facilitate payments, fulfill their tasks, and deliver goods through a secure supply chain~\cite{astarita2020review}~\cite{mollah2020blockchain}~\cite{lang2015managing}. However, the application of access control delegation using blockchain in transportation, in particular, for transportation resource allocation, is lacking in most of these proposals.

For instance, to improve transportation infrastructures,  proposal~\cite{vazquez2021towards} discusses a blockchain-based real-time ride-sharing service. It mainly focuses on efficient vehicle utilisation so that the number of vehicles on the road can be minimised, reducing traffic congestion.  In~\cite{yuan2016towards}, a blockchain-based intelligent transportation system is discussed. 
Blockchain is used to establish a decentralised autonomous platform for the transportation ecosystem. In~\cite{jiang2020intelligent}, a blockchain-based resource allocation technique is discussed. A video analysis framework is used to improve traffic safety and efficiency for the Internet of Autonomous Vehicles (IoAV) infrastructure. Blockchain is used to reduce the latency of the system. While these proposals discuss security, mobility, automation, latency, and communications issues, they do not focus on specific resource allocation (e.g., transportation credits) issue. Unlike these approaches, in our work, we devise a distributed and decentralised management of transportation credits among various organisations. This enhances the delegation issues to a more fine-grained level for the allocation of access rights. 

Proposals~\cite{gupta2019dynamic} and~\cite{gupta2020attribute} present an ABAC model for secure distribution of sensitive information of vehicles (e.g., location, time,  speed, etc.) in an intelligent transportation system. However, these proposals are built over a highly centralised component for authorisation and access right delegation. While they provide scalability to the system by enhancing attributes-based security policies, unlike our proposal, the allocation of resources that can be managed for individual entities is missing in these proposals. 

In~\cite{verdonck2013collaborative}, a collaborative supply chain framework for transportation companies is discussed. It focuses on the company's logistic functionalities, e.g., order and capacity sharing. In~\cite{darwish2018fog}, a fog-based architecture is discussed for distribution and data processing of transport network resources at the edge nodes. It helps seamless and faster responses and coordinates resources and application queries in real-time. Similar to \cite{darwish2018fog}, proposal \cite{xhafa2021allocation} present a fog-based resource allocation techniques for intelligent transportation systems. Proposal \cite{gerla2014internet} presents an `Internet of Vehicles' (IoV) concept from an energy management point of view. Based on the energy, it decides to drive customers to a particular destination. In~\cite{zhao2021blockchain}, a blockchain-based renewable energy management process is developed for electronic vehicles. It focuses on the improvement of energy efficiency at the time of charging the vehicles. A similar approach can be seen in proposal~\cite{ning2021blockchain}, but it uses a blockchain-enabled crowdsensing framework. However, these proposals do not discuss the resource allocation issue. 

Our proposed framework takes advantage of blockchain technology to overcome the limitations of the traditional approaches (e.g., RBAC, ABAC, and CapBAC) and leverage the potential of safe, secure, and fine-grained allocation of transportation credits among various entities without the need for a centralised trust management system.

\section{Conclusion and Future Work}
\label{conclusion}
In this paper, we have discussed the allocation of resources from a multi-organisational point of view supported by blockchain technology. We used access control delegation to transfer, maintain, and control the transportation credits over the employees within an organisation. The proposed framework employed blockchain smart contracts to maintain safe and secure interactions among transportation companies and organisations without the need for a trusted third-party. The use of blockchain smart contracts enables the resource exchange to occur in an open market place without an imbalance of power between the participants. it also allows multiple providers and consumers to participate in the market. This provides a flexible, open and secure system leveraging off the natureof the blockchain. 

The proof of implementation prototype showed that the proposed framework is feasible to use in city-wide transport, even potentially country-wide large-scale scenarios. HLF is an enterprise ready blockchain that can achieve a throughput of up to 10,000 transactions per second. This indicates that a large scale deployment using HLF would not have network bottlenecks and would be able to provide a seamless experience for users. One of the challenges in a large scale deployment would be how to guarantee decentralisation, i.e., how do we ensure that no single entity is ever in control of the endorsers and validators. A potential solution would be to have each transport company maintain their own endorser and validator. A user can then set endorsement policies that ensure each transport company must use at least 30\% of total nodes. This reduces the possibility that a single entity or group of entities can control the endorsing and validating stages.  A limitation of this work is that even with separate communication channels for each organisation, there is a security risk of confidential information leaking. Although this can be mitigated with private data collections and having anonymous transactions there is still the inherent risk of those transactions being broadcast. With these solutions in place, our system could easily scale up to a large production system without loss of efficiency or quality. 

In the future, we plan to conduct more experiments to assess the security level and see the feasibility of our framework with more complex access control policies. We also plan to add more application features to transport chain e.g., route planning and automatic token purchase based on patterns in the historical transit data. Each of these features would add another aspect of that can be seen in the traditional transport sector today, with the goal of creating a transparent and open transport system that is truly decentralised. We plan on assessing the effectiveness of such a transport system and highlighting the major benefits of implementing such a system.

\ifCLASSOPTIONcaptionsoff
  \newpage
\fi

\bibliographystyle{IEEEtran}
\bibliography{bare_jrnl}

\end{document}